\documentclass[11pt,a4paper]{article}
\usepackage{amsmath, amssymb}
\usepackage{latexsym, color}
\usepackage{epsfig}

\numberwithin{equation}{section}

\newtheorem{thm}{Theorem}[section]

\newtheorem{lem}[thm]{Lemma}
\newtheorem{prop}[thm]{Proposition}

\def\nm{\noalign{\medskip}}

\newcommand{\qed}{\hfill \ensuremath{\square}}
\newcommand{\ds}{\displaystyle}
\newcommand{\pf}{\noindent {\sl Proof}. \ }
\newcommand{\p}{\partial}
\newcommand{\pd}[2]{\frac {\p #1}{\p #2}}

\newcommand{\eqnref}[1]{(\ref {#1})}

\newcommand{\zbar}{\overline{z}}

\newcommand{\hatgrad}{\widehat{\nabla}}

\newcommand{\Cbb}{\mathbb{C}}

\newcommand{\Rbb}{\mathbb{R}}

\renewcommand{\div}{\mbox{div}}

\def\Ba{{\bf a}}
\def\Bb{{\bf b}}

\def\Bh{{\bf h}}

\def\Bn{{\bf n}}

\def\Bu{{\bf u}}
\def\Bv{{\bf v}}

\def\Bx{{\bf x}}
\def\By{{\bf y}}

\def\BA{{\bf A}}
\def\BB{{\bf B}}

\def\BI{{\bf I}}

\newcommand{\Ga}{\alpha}
\newcommand{\Gb}{\beta}
\newcommand{\Gd}{\delta}
\newcommand{\Ge}{\epsilon}

\newcommand{\Gvf}{\varphi}
\newcommand{\Gg}{\gamma}

\newcommand{\Gk}{\kappa}

\newcommand{\Gm}{\mu}

\newcommand{\Gs}{\sigma}

\newcommand{\GD}{\Delta}

\newcommand{\GG}{\Gamma}

\newcommand{\GO}{\Omega}

\newcommand{\beq}{\begin{equation}}
\newcommand{\eeq}{\end{equation}}
\newcommand{\ol}{\overline}

\begin{document}
\title{On coated inclusions neutral to bulk strain fields in two dimensions\thanks{\footnotesize This work is supported by the A3 Foresight Program among China, Japan, and Korea through NRF grant NRF-2014K2A2A6000567.}}

\author{Hyeonbae Kang\thanks{\footnotesize Department of Mathematics, Inha University, Incheon
22212, S. Korea (hbkang@inha.ac.kr).} }

\date{}
\maketitle

\begin{abstract}
The neutral inclusion problem in two dimensional isotropic elasticity is considered. The neutral inclusion, when inserted in a matrix having a uniform applied field, does not disturb the field outside the
inclusion. The inclusion consists of the core and shell of arbitrary shapes, and their elasticity tensors are isotropic.  We show that if the coated inclusion is neutral to a uniform bulk field, then the core and shell must be concentric disks, provided that the shear and bulk moduli satisfy certain conditions.
\end{abstract}


\section{Introduction}

Some inclusions, when inserted in a matrix having a uniform field, do not disturb the field outside the
inclusion. Such inclusions are called neutral inclusions (to the given field). A typical neutral inclusion consists of a core coated by a shell having the material property different from that of the core.

It is easy to construct neutral inclusions of circular shapes in the context of conductivity (or anti-plane elasticity). Let $D=\{\, |\Bx|< r_1\,\}$ and $\GO=\{\, |\Bx|< r_2\,\}$ ($r_1<r_2$) so that $D$ is the core and $\GO \setminus D$ is the shell. The conductivity is $\Gs_c$ in the core, $\Gs_s$ in the shell, and $\Gs_m$ in the matrix ($\Rbb^2 \setminus \GO$). So the conductivity distribution is given by
$$
\Gs = \Gs_c \chi(D) + \Gs_s \chi(\GO \setminus D) + \Gs_m \chi(\Rbb^2 \setminus \GO)
$$
where $\chi$ is the characteristic function. If $\Gs_c$, $\Gs_s$ and $\Gs_m$ satisfy the relation
\beq\label{neutral_cond}
r_2^2 (\Gs_s + \Gs_c)(\Gs_m - \Gs_s) - r_1^2 (\Gs_s - \Gs_c)(\Gs_m + \Gs_s) =0,
\eeq
then $\GO$ is neutral to uniform fields. In other words, for any constant vector $\Ba$, the solution $u$ to the problem
$$
\left\{
\begin{array}{ll}
\nabla \cdot \Gs \nabla u = 0 \quad &\mbox{in } \Rbb^2, \\
u(\Bx)- \Ba \cdot \Bx = O(|\Bx|^{-1}) \quad &\mbox{as } |\Bx| \to \infty
\end{array}
\right.
$$
satisfies $u(\Bx)-\Ba \cdot \Bx=0$ in $\Rbb^2 \setminus \GO$.

Much interest in neutral inclusions was aroused by the work of Hashin \cite{hashine, hashine2}, where it is shown that since insertion of neutral inclusions does not perturb the outside uniform field, the effective conductivity of the assemblage filled with neutral inclusions of many different scales is $\Gs_m$ satisfying \eqnref{neutral_cond}. It is also proved that this effective conductivity is a bound of the Hashin-Shtrikman bounds on the effective conductivity of arbitrary two phase composites. We refer to a book of Milton \cite{milton} for development on neutral inclusions in relation to theory of composites.

Another interest in neutral inclusions has aroused in relation to the invisibility cloaking by transformation optics. In this regard, we first observe that in general the solution $u$ to
\beq\label{transcond}
\left\{
\begin{array}{ll}
\nabla \cdot \Gs \nabla u = 0 \quad &\mbox{in } \Rbb^2, \\
u(\Bx)- h(\Bx) = O(|\Bx|^{-1}) \quad &\mbox{as } |\Bx| \to \infty
\end{array}
\right.
\eeq
for a given harmonic function $h$ satisfies $u(\Bx)-h(\Bx)=O(|\Bx|^{-1})$ as $|\Bx| \to \infty$. But, if the inclusion is neutral to all uniform fields, then the linear part of $h$ is unperturbed and one can show using multi-polar expansions that $u(\Bx)-h(\Bx)=O(|\Bx|^{-2})$ as $|\Bx| \to \infty$ for any $h$ (not necessarily linear). Recently, Ammari {\it et al} \cite{AKLL1} extend the idea of neutral inclusions to construct multi-coated circular structures which are neutral not only to uniform fields but also to fields of higher order, so that the solution $u$ to \eqnref{transcond} satisfies $u(\Bx)-h(\Bx)=O(|\Bx|^{-N})$ as $|\Bx| \to \infty$ for any given $N$ and any $h$ (such structures are called GPT vanishing structures). Such structures have a strong connection to the cloaking by transformation optics. The transformation optics proposed by Pendry {\it et al} \cite{pendry} transforms a punctured disk (or a sphere) to an annulus to achieve perfect cloaking. The same transform was used to show non-uniqueness of the Calder\'on's problem by Greenleaf {\it et al} \cite{glu}. Kohn {\it et al} \cite{KSVW-IP-08} showed that if one transforms a disk with small hole, then one can avoid singularities of the conductivity which occur on the inner boundary of the annulus and achieve near-cloaking instead of the perfect cloaking. In \cite{AKLL1} it is shown that if we coat the hole by multiple layers so that the structure becomes neutral to fields of higher order (and transform the structure), then the near-cloaking effect is dramatically improved.

All above mentioned neutral inclusions have circular shapes and it is of interest to consider neutral inclusions of arbitrary shapes. For a given core of arbitrary shape, the shape of the outer boundary of the shell has been constructed by Milton \& Serkov \cite{MS} so that the coated inclusion is neutral to a single uniform field. This is done when the conductivity $\Gs_c$ of the core is either $0$ or $\infty$. See \cite{JM} for an extension to the case when $\Gs_c$ is finite. It is also proved in \cite{MS} that if an inclusion is neutral to all uniform field (or equivalently, to two linearly independent uniform fields), then the inclusion is concentric disks (confocal ellipses if the conductivity of the matrix is anisotropic), when $\Gs_c$ is $0$ or $\infty$. In recent paper \cite{KL}, Kang and Lee proved that this is the case even when $\Gs_c$ is finite. See also \cite{KLS} for an extension to three dimensions.

In this paper the problem of neutral inclusions in two dimensional linear isotropic elasticity is considered. Let the shear and bulk moduli of the core, the shell, and the matrix be $(\mu_c,\Gk_c)$, $(\mu_s,\Gk_s)$, and $(\mu_m,\Gk_m)$, respectively, and let $\mu$ and $\Gk$ denote their distributions in $\Rbb^2$. Define the elasticity tensor $\Cbb=(C_{ijkl})$ by
\beq
C_{ijkl} =
(\Gk-\mu) \delta_{ij}\delta_{k\ell} + \Gm (\delta_{ik}\delta_{j\ell}+\delta_{i\ell}\delta_{jk}), \quad i,j,k,l=1,2 \, ,
\eeq
where $\Gd_{ij}$ is the Kronecker's delta. Let $\Bh(\Bx)= \Bx$, whose gradient represents the bulk strain field,
and consider the following interface problem:
\begin{equation}\label{main-eqn}
\left\{
\begin{array}{ll}
\div \, \Cbb \hatgrad \Bu=0 \quad & \mbox{in } \Rbb^2,  \\
\nm \Bu(\Bx)- \Bh(\Bx)=O(|\Bx|^{-1}) \quad & \mbox{as
}|\Bx|\rightarrow \infty,
\end{array}
\right.
\end{equation}
where $\hatgrad \Bu$ is the symmetric gradient (or the strain tensor), {\it i.e.},
$$
\hatgrad \Bu:= \frac{1}{2} (\nabla \Bu + (\nabla \Bu)^{T}) \quad\text{($T$ for transpose)}.
$$
The inclusion is neutral to the (strain) field $\nabla \Bh$ if the solution $\Bu$ to \eqnref{main-eqn} satisfies $\Bu(\Bx)- \Bh(\Bx)=0$ in $\Rbb^2 \setminus \GO$. Inclusions neutral to the bulk field was found using the exact effective bulk modulus of the assemblage of coated disks which was derived by Hashin and Rosen \cite{HR}. The purpose of this paper is to prove that concentric disks are the only coated inclusions neutral to bulk fields under some conditions on the shear and bulk moduli.

The following is the main theorem of this paper.

\begin{thm}\label{thm:bulk}
Let $\GO$ and $D$ be bounded simply connected domains in $\Rbb^2$ with Lipschitz boundaries such that $\ol{D} \subset \GO$. Suppose that
\beq\label{hypo}
\Gm_c \neq \Gm_s, \quad \Gk_m \neq \Gk_s,  \quad\text{and}\quad \Gk_c < 2 \Gk_s + \Gm_s.
\eeq
If $(\GO, D)$ is neutral to the bulk field, or equivalently, if the solution $\Bu$ to \eqnref{main-eqn} with $\Bh(\Bx)=\Bx$ satisfies $\Bu(\Bx)- \Bx=0$ in $\Rbb^2 \setminus \GO$, then $D$ and $\GO$ are concentric disks.
\end{thm}

The conditions in \eqnref{hypo} are required to show that the solution is linear in the core. The first two conditions seem natural because the elasticity properties of the core, the shell, and the matrix must be different. However, we don't know if the third condition is necessary.

It is worth mentioning that inclusions consisting of the concentric disks are not neutral to shear fields: for example, if $\Bh(\Bx)= (y,x)^T$, then $u(\Bx)-\Bh(\Bx)$ has a term of order $|\Bx|^{-1}$ and a term of order $|\Bx|^{-3}$ as $|\Bx| \to \infty$, and it is not possible to make both terms vanish. Christensen and Lo \cite{CL} constructed circular inclusions such that the term of order $|\Bx|^{-1}$ vanishes and derived an effective transverse shear modulus of the assemblage of coated disks. It is interesting to construct coated inclusions neutral to shear fields or to prove non-existence of such inclusions.

The rest of the paper is organized as follows: In the next section we show that if $(\GO, D)$ is neutral to the bulk field, then $\nabla \Bu$ is symmetric and $\div \Bu$ is constant in the shell. The main theorem is proved in section \ref{sec:bulk} by showing that $\Bu$ is linear in the core. To do so we use a complex representation of the displacement vector.

\section{Properties of the solution in the shell}

In this section we prove the following proposition. We emphasize that \eqnref{hypo} is not required for this proposition.
\begin{prop}\label{prop:shell}
Let $\GO$ and $D$ be bounded simply connected domains in $\Rbb^2$ with Lipschitz boundaries such that $\ol{D} \subset \GO$. If $(\GO,D)$ is neutral to the bulk field, then the solution $\Bu$ to \eqnref{main-eqn} satisfies the following:
\begin{itemize}
\item[(i)] $\GD \Bu=0$, or equivalently $\text{div} \,\Bu=\text{constant}$ in $\GO \setminus \ol{D}$.
\item[(ii)] There is a function $\chi$ such that $\Bu = \nabla \chi$ in $\Rbb^2 \setminus \ol{D}$. In particular, $\nabla \Bu$ is symmetric in $\GO \setminus \ol{D}$.
\end{itemize}
\end{prop}

To prove Proposition \ref{prop:shell}, we need some preparartion. The Kelvin matrix ${\bf \GG (\Bx)} = ( \GG_{ij}(\Bx) )_{i,j=1}^2$ of the
fundamental solution  to the Lam{\'e} operator $\div \, \Cbb \hatgrad$ in two dimensions is given by
\beq \label{Kelvin}
 \GG_{ij} (\Bx) : =  \frac{\Ga_1}{2 \pi} \Gd_{ij} \log |\Bx| -  \frac{\Ga_2}{2 \pi}
 \frac{x_i x_j}{|\Bx|^2}, \quad  \Bx \neq 0\;,
\eeq
where
 \beq \label{ab}
 \Ga_1= \frac{1}{2} \left ( \frac{1}{\mu} +
 \frac{1}{\mu + \Gk} \right ) \quad\mbox{and}\quad
 \Ga_2=
 \frac{1}{2} \left ( \frac{1}{\mu} - \frac{1}{\mu + \Gk}
 \right )\;.
 \eeq
A straight-forward computation shows that
\beq\label{divGG}
\div_\By \,{\bf \GG}(\Bx-\By) = \frac{\Ga_2 - \Ga_1}{2\pi} \nabla_\Bx \log|\Bx-\By| = - \frac{1}{2\pi(\mu+\Gk)} \nabla_\Bx \log|\Bx-\By| \, .
\eeq
In particular, we have
\beq\label{divGG2}
\int_{\GO} \mbox{\rm div}_\By \,{\bf \GG}(\Bx-\By) d\By = - \frac{1}{2\pi(\mu+\Gk)} \nabla \int_{\GO} \log|\Bx-\By| d\By.
\eeq
Since
$$
\frac{1}{2\pi} \GD \int_{\GO} \log|\Bx-\By| d\By = \begin{cases}
\ds 1 &\quad \mbox{if } \Bx \in \GO, \\
0 &\quad \mbox{if } \Bx \in \Rbb^2 \setminus \overline{\GO},
\end{cases}
$$
we have
\beq\label{divdiv}
\mbox{\rm div} \, \int_{\GO} \mbox{\rm div}_\By \, {\bf \GG}(\Bx-\By) d\By = \begin{cases}
\ds - \frac{1}{\mu + \Gk} &\quad \mbox{if } \Bx \in \GO, \\
0 &\quad \mbox{if } \Bx \in \Rbb^2 \setminus \overline{\GO}.
\end{cases}
\eeq
We also have
\beq\label{curldiv}
\mbox{\rm rot} \, \int_{\GO} \mbox{\rm div}_\By \, {\bf \GG}(\Bx-\By) d\By = 0 \, .
\eeq

\medskip

\noindent{\sl Proof of Proposition \ref{prop:shell}}.
Suppose that $(\GO, D)$ is neutral to the bulk field. Then the following over-determined problem admits a solution:
\beq\label{trans}
\left\{
\begin{array}{ll}
\nabla \cdot (\Cbb \hatgrad \Bu) = 0 \quad &\mbox{in } \GO, \\
\Bu(\Bx)=\Bx, \ \ (\Cbb_s \hatgrad \Bu) \Bn = (\Cbb_m \BI) \Bn \quad &\mbox{on } \p\GO.
\end{array}
\right.
\eeq
Here and throughout this paper, $\Bn$ denotes the outward normal to $\p\GO$ (and $\p D$). Let $\Bu_c$ and $\Bu_s$ denote the solution on $D$ and $\GO \setminus \ol{D}$, respectively. Then the transmission condition along $\p D$ is given by
\beq\label{trans2}
\Bu_c=\Bu_s \quad\text{and}\quad (\Cbb_c \hatgrad \Bu_c) \Bn=(\Cbb_s \hatgrad \Bu_s) \Bn \quad\text{on } \p D.
\eeq

Let $\Bv$ be a smooth vector field in $\GO$. Then we have
$$
\int_{\p\GO} (\Cbb_s \hatgrad \Bu) \Bn \cdot \Bv \, d\Gs = \int_{\GO} \Cbb \hatgrad \Bu : \hatgrad \Bv \, d\By.
$$
Here and afterwards, $\BA:\BB$ denotes the contraction of two matrices $\BA$ and $\BB$,
{\it i.e.}, $\BA:\BB=\sum a_{ij}b_{ij}=\textrm{tr}(\BA^{T}\BB)$. On the other hand, we have from the Neumann boundary condition in \eqnref{trans}
$$
\int_{\p\GO} (\Cbb \hatgrad \Bu) \Bn \cdot \Bv \, d\Gs = \int_{\GO} \Cbb_m \BI : \hatgrad \Bv \, d\By.
$$
So, we have
\beq\label{idone}
\int_{\GO\setminus D} \Cbb_s \hatgrad \Bu : \hatgrad \Bv \, d\By + \int_{D} \Cbb_c \hatgrad \Bu : \hatgrad \Bv \, d\By
= \int_{\GO} \Cbb_m \BI : \hatgrad \Bv \, d\By.
\eeq

Using the Dirichlet boundary condition in \eqnref{trans} we have for any elasticity tensor $\Cbb_0$
$$
\int_{\p\GO} \Bu \cdot (\Cbb_0 \hatgrad \Bv) \Bn \, d\Gs = \int_{\GO} \Cbb_0 \hatgrad \Bu : \hatgrad \Bv \, d\By + \int_{\GO} \Bu \cdot \div (\Cbb_0 \hatgrad \Bv) \, d\By
$$
and
$$
\int_{\p\GO} \Bu \cdot (\Cbb_0 \hatgrad \Bv) \Bn \, d\Gs = \int_{\p\GO} \By \cdot (\Cbb_0 \hatgrad \Bv) \Bn \, d\Gs
= \int_{\GO} \Cbb_0 \BI : \hatgrad \Bv \, d\By + \int_{\GO} \By \cdot \div (\Cbb_0 \hatgrad \Bv) \, d\By.
$$
Thus we have
\beq\label{idtwo}
\int_{\GO} \Cbb_0 \hatgrad \Bu : \hatgrad \Bv \, d\By + \int_{\GO} \Bu \cdot \div (\Cbb_0 \hatgrad \Bv) \, d\By
= \int_{\GO} \Cbb_0 \BI : \hatgrad \Bv \, d\By + \int_{\GO} \By \cdot \div (\Cbb_0 \hatgrad \Bv) \, d\By.
\eeq
Subtracting \eqnref{idtwo} with $\Cbb_0=\Cbb_s$ from \eqnref{idone} we obtain
\begin{align}
& \int_{D} (\Cbb_c - \Cbb_s) \hatgrad \Bu : \hatgrad \Bv \, d\By - \int_{\GO} \Bu \cdot \div (\Cbb_s \hatgrad \Bv) \, d\By \nonumber \\
& = \int_{\GO} (\Cbb_m - \Cbb_s) \BI : \hatgrad \Bv \, d\By - \int_{\GO} \By \cdot \div (\Cbb_s \hatgrad \Bv) \, d\By. \label{idthree}
\end{align}

Let ${\bf \GG}^s$ and ${\bf \GG}^c$ be the Kelvin matrices for $\div\, \Cbb_s \hatgrad$ and $\div\, \Cbb_c \hatgrad$, respectively. For $\Bx \in \GO$, let $\Bv(\By)$ be a column of ${\bf \GG}^s (\Bx-\By)$. Then we may apply the same argument of integration by parts (over $\GO$ with an $\Ge$ ball around $\Bx$ deleted) as above and obtain from \eqnref{idthree} the following representation of the solution:
\begin{align}
\Bu(\Bx) &= \Bx + \int_{D} (\Cbb_c - \Cbb_s) \hatgrad \Bu(\By) : \hatgrad {\bf \GG}^s (\Bx-\By) \, d\By \nonumber \\
 & \quad + \int_{\GO} (\Cbb_s - \Cbb_m) \BI : \hatgrad {\bf \GG}^s (\Bx-\By) \, d\By, \quad \Bx \in \GO . \label{repone}
\end{align}

Since $(\Cbb_m - \Cbb_s) \BI : \hatgrad \Bv = 2(\Gk_m - \Gk_s) \div\, \Bv$,
the identity \eqnref{idthree} takes the form
\begin{align}
&\int_{D} (\Cbb_c - \Cbb_s) \hatgrad \Bu : \hatgrad \Bv \, d\By - \int_{\GO} \Bu \cdot \div (\Cbb_s \hatgrad \Bv) \, d\By \nonumber \\
&= 2(\Gk_m - \Gk_s) \int_{\GO} \div\, \Bv \, d\By - \int_{\GO} \By \cdot \div (\Cbb_s \hatgrad \Bv) \, d\By, \label{idbulkone}
\end{align}
One can also see from \eqnref{divGG2} that the representation formula \eqnref{repone} takes the form
\begin{align}
\Bu(\Bx) &=  \Bx + \int_{D} (\Cbb_c - \Cbb_s) \hatgrad \Bu(\By) : \hatgrad_\By {\bf \GG}^s (\Bx-\By) \, d\By \nonumber \\
 & \quad - \frac{\Gk_m - \Gk_s}{\pi(\mu_s+\Gk_s)} \nabla \int_{\GO} \log|\Bx-\By| \, d\By, \quad \Bx \in \GO . \label{repbulkone}
\end{align}

Let ${\bf \GG}^{s,j}$ be the $j$-th column of ${\bf \GG}^s$. Let $\Bx \in \Rbb^2 \setminus \overline{\GO}$. Substitute $\Bv_j(\By) := \pd{}{x_j} {\bf \GG}^{s,j} (\Bx-\By)$ for $\Bv$ in \eqnref{idbulkone} and add the identities for $j=1,2$. Note that $\div (\Cbb_s \hatgrad \Bv_j)=0$ in $\GO$ since $\Bx \notin \GO$. We infer from \eqnref{divdiv} that
$$
\sum_{j=1}^2 \int_\GO \div\, \Bv_j = \sum_{j=1}^2 \pd{}{x_j} \int_\GO \div_\By \, {\bf \GG}^{s,j} (\Bx-\By) \, d\By = 0.
$$
It then follows from \eqnref{idbulkone} that
\beq\label{zeroD}
\sum_{j=1}^2 \pd{}{x_j} \int_{D} (\Cbb_c - \Cbb_s) \hatgrad \Bu(\By) : \hatgrad {\bf \GG}^{s,j} (\Bx-\By) \, d\By =0, \quad \Bx \in \Rbb^2 \setminus \overline{\GO}.
\eeq
Observe that the left-hand side in the above is a real analytic function in $\Rbb^2 \setminus \ol{D}$. So, by unique continuation \eqnref{zeroD} holds for all $\Bx \in \Rbb^2 \setminus \overline{D}$. We then infer from \eqnref{divdiv} and \eqnref{repbulkone} that
\beq\label{divconst}
\div\, \Bu = \Ga \quad\mbox{in } \GO \setminus \ol{D},
\eeq
where $\Ga$ is the constant given by
\beq\label{Galpha}
\Ga = 2- \frac{2(\Gk_m - \Gk_s)}{\mu_s + \Gk_s} .
\eeq
Since $\div (\Cbb_s \hatgrad \Bu)= \mu_s \GD \Bu + \Gk_s \nabla \div \Bu=0$, we also have
\beq
\GD \Bu =0 \quad\mbox{in } \GO \setminus D.
\eeq

We now prove (ii). Let $\Bx \in \Rbb^2 \setminus \ol{\GO}$ and substitute $\pd{}{x_2} {\bf \GG}^{s,1} (\Bx-\By)$ for $\Bv$  in \eqnref{idbulkone} to obtain from \eqnref{divGG2} that
\begin{align*}
\pd{}{x_2} \int_{D} (\Cbb_c - \Cbb_s) \hatgrad \Bu(\By) : \hatgrad {\bf \GG}^{s,1} (\Bx-\By) \, d\By
& = 2(\Gk_m - \Gk_s) \pd{}{x_2} \int_D \div_\By \, {\bf \GG}^{s,1} (\Bx-\By) \, d\By \\
& = - \frac{(\Gk_m - \Gk_s)}{\pi(\mu_s+\Gk_s)} \frac{\p^2}{\p x_1 \p x_2} \int_{D} \log|\Bx-\By| \, d\By \, .
\end{align*}
By substituting $\pd{}{x_1} {\bf \GG}^{s,2} (\Bx-\By)$ for $\Bv$  in \eqnref{idbulkone}, we also obtain
\begin{align*}
\pd{}{x_1} \int_{D} (\Cbb_c - \Cbb_s) \hatgrad \Bu(\By) : \hatgrad {\bf \GG}^{s,2} (\Bx-\By) \, d\By
= - \frac{(\Gk_m - \Gk_s)}{\pi(\mu_s+\Gk_s)} \frac{\p^2}{\p x_1 \p x_2} \int_{D} \log|\Bx-\By| \, d\By \, .
\end{align*}
So, we have using unique continuation again that
\begin{align*}
& \pd{}{x_2} \int_{D} (\Cbb_c - \Cbb_s) \hatgrad \Bu(\By) : \hatgrad {\bf \GG}^{s,1} (\Bx-\By) \, d\By \\
& = \pd{}{x_1} \int_{D} (\Cbb_c - \Cbb_s) \hatgrad \Bu(\By) : \hatgrad {\bf \GG}^{s,2} (\Bx-\By) \, d\By, \quad \Bx \in \Rbb^2 \setminus \ol{D} .
\end{align*}
Since $\int_{D} (\Cbb_c - \Cbb_s) \hatgrad \Bu(\By) : \hatgrad {\bf \GG}^{s} (\Bx-\By) \, d\By \to 0$ as $|\Bx| \to \infty$, we infer that there is $\chi_1$ such that
$$
\int_{D} (\Cbb_c - \Cbb_s) \hatgrad \Bu(\By) : \hatgrad {\bf \GG}^{s} (\Bx-\By) \, d\By = \nabla \chi_1(\Bx).
$$
Let
$$
\chi(\Bx):= \frac{1}{2} |\Bx|^2  + \chi_1(\Bx) - \frac{2(\Gk_m - \Gk_s)}{\mu_s+\Gk_s} \int_{\GO} \log|\Bx-\By| \, d\By.
$$
Then (ii) is proved. \qed

\section{Neutral Inclusions to the bulk field}\label{sec:bulk}

\subsection{Complex representation of the solution and a lemma}

Let $\Bu= (u_1, u_2)^T$ be the solution to \eqnref{main-eqn}. There are functions $\Gvf$ and $\psi$ which are analytic in $D$, $\GO\setminus \ol{D}$, and $\Cbb \setminus \ol{D}$, separately, such that
\beq\label{complexform}
u_1 + iu_2 = \frac{1}{2\mu}\left( k \varphi(z) - z \ol{\varphi'(z)}- \ol{\psi(z)}\right),
\eeq
where
\beq\label{kkrel}
k = 1+ \frac{2\mu}{\Gk}.
\eeq
See for example \cite{book2, musk}. Conversely, $\Bu= (u_1, u_2)^T$ of the form \eqnref{complexform} with $k >1$ for a pair of analytic functions $\Gvf$ and $\psi$ in $D$ is a solution in $D$ of the Lam\'e system determined by the shear modulus $\Gm$ and the bulk modulus $\Gk = 2\Gm/(k-1)$.

We denote $\Gvf$ and $\psi$ by $\Gvf_c$ and $\psi_c$ in the core, $\Gvf_s$ and $\psi_s$ in the shell, and $\Gvf_m$ and $\psi_m$ in the matrix. Then the transmission conditions \eqnref{trans} and \eqnref{trans2} along the interfaces $\p D$ and $\p \GO$ take the following forms: along $\p D$,
\begin{align*}
\ds \frac{1}{2\mu_s} \bigg( k_s \Gvf_s(z) - z \ol{\Gvf_s'(z)} - \ol{\psi_s(z)} \bigg )
& = \ds \frac{1}{2\mu_c} \bigg ( k_c \Gvf_c(z) - z \ol{\Gvf_c'(z)} -
 \ol{\psi_c(z)} \bigg ),  \\
 d(\Gvf_s(z) + z \ol{\Gvf_s'(z)} + \ol{\psi_s(z)}) & =
 d(\Gvf_c(z) + z \ol{\Gvf_c'(z)} + \ol{\psi_c(z)})  ,
\end{align*}
and similar conditions on $\p\GO$, where $d$ is the exterior differential. The first condition is the continuity of the displacement and the second one is that of the traction. Using complex notation $dz=dx+i dy$ and $d \zbar=dx-idy$, the exterior differential is given by
$$
df = \pd{f}{z} dz+ \pd{f}{\zbar} d\zbar.
$$
It is convenient to use notation
\beq\label{U(z)}
U(z):= u_1 + iu_2 = \frac{1}{2\mu}\left( k \varphi(z) - z \ol{\varphi'(z)}- \ol{\psi(z)}\right),
\eeq
and
\beq\label{DU(z)}
DU(z) := d(\Gvf + z \ol{\Gvf'} + \ol{\psi})= (\Gvf'+ \ol{\Gvf'}) dz + (z \ol{\Gvf''} + \ol{\psi'})d\zbar.
\eeq
Then the transmission conditions read
\beq\label{transD}
U_c=U_s, \quad DU_c=DU_s \quad\mbox{on } \p D,
\eeq
and
\beq\label{transOm}
U_m=U_s, \quad DU_m=DU_s \quad\mbox{on } \p \GO.
\eeq

The proofs in the subsequent subsection use the following lemma, which may be well-known. We include a short proof for readers' sake.
\begin{lem}\label{lemma}
Let $D$ be a simply connected bounded domain with the Lipschitz boundary, and let $g$ be a square integrable function on $\p \GO$. If
\beq\label{ortho}
\int_{\p D} g(z) f'(z) dz=0
\eeq
for any function $f$ analytic in a neighborhood of $\ol{D}$, then there is an analytic function $G$ in $D$ such that $G=g$ on $\p D$.
\end{lem}

\pf Define the Cauchy transform by
$$
C[g](w):= \frac{1}{2\pi i} \int_{\p D} \frac{g(z)}{z-w} dz, \quad w \in \Cbb \setminus \p D.
$$
Then by Plemelj's jump formula (see \cite{musk}), we have
$$
g(w)= C[g]|_{-}(w) - C[g]|_{+}(w), \quad w \in \p D,
$$
where $C[g]|_{-}$ and $C[g]|_{+}$ denote the limits from inside and outside of $ D$, respectively. Since $ D$ is simply connected, $f(z)=\log (z-w)$ is well-defined and analytic in a neighborhood of $\ol{D}$ if $w \notin \ol{D}$. So, $C[g](w)=0$ if $w \notin \ol{ D}$ by \eqnref{ortho}. Thus, we have
$$
g(w)= C[g]|_{-}(w), \quad w \in \p D.
$$
So, $G:= C[g]$ in $D$ is the desired analytic function. \qed

\subsection{Proof of Theorem \ref{thm:bulk}}\label{sec:proof}

Let us prove the following proposition first.
\begin{prop}\label{prop:core}
Let $\GO$ and $D$ be bounded simply connected domains in $\Rbb^2$ with Lipschitz boundaries such that $\ol{D} \subset \GO$, and assume that \eqnref{hypo} holds. If $(\GO, D)$ is neutral to the bulk field, then the solution $\Bu$ to \eqnref{main-eqn} is linear in $D$ and of the form
\beq\label{axb}
\Bu(\Bx) = a \Bx + \Bb
\eeq
for a constant $a$ and a constant vector $\Bb$.
\end{prop}

\pf
Let $\Bu$ be the solution to \eqnref{main-eqn} when $\Bh(\Bx)=\Bx$, and $U$ be defined by \eqnref{U(z)}. Since $(\GO, D)$ is neutral to the bulk field, $\Bu(\Bx)=\Bx$ in $\Rbb^2 \setminus \ol{\GO}$, and hence we have
$$
U_m(z)=z, \quad \Gvf_m(z)=\Gk_m z, \quad \psi_m(z)=0, \quad DU_m(z) =2 \Gk_m dz.
$$
Moreover, Proposition \ref{prop:shell} implies that
\beq\label{shell}
\Gvf_s(z) = \Gb z + \text{constant}, \quad z \in \GO \setminus D,
\eeq
where $\Gb$ is a real constant. In fact, we see from Proposition \ref{prop:shell} that
$$
\frac{\p}{\p z} U_s = \frac{1}{2} \left( \frac{\p}{\p x_1} - i \frac{\p}{\p x_2} \right) (u_1+ i u_2) = \frac{1}{2} \left (\div \Bu + i \left( \pd{u_2}{x_1} - \pd{u_1}{x_2} \right) \right)=\frac{\Ga}{2},
$$
where $\Ga$ is the constant in \eqnref{Galpha}. Thus we have
$$
\frac{\Ga}{2}= \frac{\p}{\p z} U_s (z) = \frac{1}{2\mu_s} \left( k_s \Gvf_s'(z) - \ol{\Gvf_s'(z)} \right),
$$
which implies that $\Gvf_s'(z)= \Gb= \Gk_s \Ga/2$ by \eqnref{kkrel}. One can see from \eqnref{Galpha} that
\beq\label{kmbeta}
\Gk_m - \Gb = \frac{(\Gk_m-\Gk_s)(2\Gk_s+\Gm_s)}{\Gk_s + \mu_s}.
\eeq

Let $f$ and $g$ be functions analytic on $\ol{\GO}$, and let
$F(z)=f(z)+\ol{g(z)}$.
We have from the first identity in \eqnref{transOm}, Cauchy's theorem and Stokes' theorem that
$$
\int_{\p\GO}U_s dF =\int_{\p\GO}U_m dF= \int_{\p\GO}z \ol{g'}\,d \zbar = - \int_{\GO} \ol{g'}\,dm \,,
$$
where $dm:= d\zbar\wedge dz$. We also have from Stokes' theorem that
\begin{align}
\int_{\p \GO}U_s dF &=\int_{\GO}d(U dF)
= \int_{\GO} \left[ \frac{\p}{\p\zbar} \left(U f' \right) - \frac{\p}{\p z} \left(U \ol{g'} \right) \right] \, dm \nonumber \\
& =- \int_{\GO} \frac{1}{2\mu} \left[ \left(z\ol{\Gvf''}+\ol{\psi'}\right)
f' + (k \Gvf'- \ol{\Gvf'}) \ol{g'} \right] \,dm\,. \label{stokes1}
\end{align}
So we have
\begin{align*}
\int_{\GO}\ol{g'}\,dm
&= \frac{1}{2\mu_s} \int_{\GO\setminus \ol{D}} \left[ \left(z\ol{\Gvf''_s}+\ol{\psi'_s}\right)
f' + (k_s \Gvf_s'- \ol{\Gvf_s'}) \ol{g'} \right] \,dm \\
& \quad + \frac{1}{2\mu_c}\int_{D} \left[ \left(z\ol{\Gvf''_c}+\ol{\psi'_c}\right)
f' + (k_c \Gvf_c'- \ol{\Gvf_c'})  \ol{g'} \right] \,dm \, .
\end{align*}
It then follows from \eqnref{kkrel} and \eqnref{shell} that
\begin{align*}
& \int_{\GO}\ol{g'}\,dm - \frac{\Gb}{\Gk_s} \int_{\GO\setminus \ol{D}}  \ol{g'} \,dm \\
&= \frac{1}{2\mu_s} \int_{\GO\setminus \ol{D}} \ol{\psi'_s} f'  \,dm + \frac{1}{2\mu_c} \int_{D} \left( z\ol{\Gvf''_c}+\ol{\psi'_c} \right)
f' \,dm + \frac{1}{2\mu_c} \int_{D} (k_c \Gvf_c'- \ol{\Gvf_c'})  \ol{g'} \,dm  \, .
\end{align*}
Since $f$ and $g$ are arbitrary, we have
\beq\label{e:genecon1}
\frac{1}{\mu_s} \int_{\GO\setminus \ol{D}} \ol{\psi'_s} f'  \,dm +
\frac{1}{\mu_c}\int_{D}\left(z\ol{\Gvf''_c}+\ol{\psi'_c}\right)f' \,dm = 0 \, ,
\eeq
and
\beq\label{e:genecon11}
\frac{1}{2\mu_c} \int_{D} (k_c \Gvf_c'- \ol{\Gvf_c'})  \ol{g'} \,dm  =  \int_{\GO}\ol{g'}\,dm - \frac{\Gb}{\Gk_s} \int_{\GO\setminus \ol{D}}  \ol{g'} \,dm \, .
\eeq

Similarly, we have from the second identity in \eqnref{transOm}
$$
\int_{\p\GO} F DU_s =\int_{\p\GO} F DU_m  = 2\Gk_m \int_{\p\GO} \ol{g} \,dz = 2\Gk_m \int_{\GO} \ol{g'} \,dm \, ,
$$
and hence
\begin{align*}
2\Gk_m \int_{\GO} \ol{g'} \,dm &=\int_{\GO} d(F DU) \\
&=\int_{\GO}
\left[ \frac{\p}{\p\zbar} \left( (f+\ol{g}) (\Gvf'+\ol{\Gvf'}) \right)
-\frac{\p}{\p z} \left( (f+\ol{g}) (z\ol{\Gvf''}+\ol{\psi'}) \right) \right] \,dm \\
&=\int_{\GO}
\left[ \ol{g'} (\Gvf'+\ol{\Gvf'}) - f' \left(z\ol{\Gvf''}+\ol{\psi'}\right) \right] dm \,.
\end{align*}
Since $f$ and $g$ are arbitrary and \eqnref{shell} holds, we obtain

\beq\label{e:genecon2}
\int_{\GO\setminus \ol{D}} \ol{\psi'_s} f' dm + \int_{D}\left(z\ol{\Gvf''_c}+\ol{\psi'_c}\right) f' \,dm =0\, ,
\eeq
and
\beq\label{e:genecon22}
\int_{D} (\Gvf_c'+\ol{\Gvf_c'}) \ol{g'} dm = 2\Gk_m \int_{\GO} \ol{g'} \,dm - 2\Gb \int_{\GO\setminus\ol{D}}
\ol{g'} dm \, .
\eeq

Since $\mu_s\neq \mu_c$ by the assumption \eqnref{hypo}, we infer from \eqnref{e:genecon1} and \eqnref{e:genecon2} that
$$
\int_{D}\left(z\ol{\Gvf''_c}+\ol{\psi'_c}\right) f' \,dm = 0 \,,
$$
or equivalently,
\beq
\int_{\p D}\left(z\ol{\Gvf'_c}+\ol{\psi_c}\right) f' \,dz = 0 \,.  \label{e:genecon3}
\eeq
Note that \eqnref{e:genecon3} holds for all functions $f$ analytic in $\ol{\GO}$. However, one can infer using Runge's approximation theorem that it holds for all $f$ analytic in a neighborhood of $\ol{D}$. So, by Lemma \ref{lemma}, there is an analytic function $\eta_1$ in $D$ such that
\beq\label{etaone}
z\ol{\Gvf'_c}+\ol{\psi_c} = \eta_1 \quad\text{on } \p D.
\eeq

On the other hand, \eqnref{e:genecon11} can be rewritten as
\beq\label{identityone}
\frac{1}{2\mu_c} \int_{D} (k \Gvf_c'- \ol{\Gvf_c'} -\frac{2\Gb \mu_c}{\Gk_s} )  \ol{g'} \,dm  =  \left( 1 - \frac{\Gb}{\Gk_s} \right) \int_{\GO}\ol{g'}\,dm \, ,
\eeq
while \eqnref{e:genecon22} as
\beq\label{identitytwo}
\int_{D} (\Gvf_c' + \ol{\Gvf_c'} -2\Gb) \ol{g'} dm = 2(\Gk_m -\Gb) \int_{\GO} \ol{g'} \,dm \, .
\eeq

Since $\Gk_m -\Gb \neq 0$ by the assumption \eqnref{hypo} and \eqnref{kmbeta}, we see from \eqnref{identityone} and \eqnref{identitytwo} that
$$
\int_{D} (k_c \Gvf_c'- \ol{\Gvf_c'} -\frac{2\Gb \mu_c}{\Gk_s} )  \ol{g'} \,dm
- \Gg \int_{D} (\Gvf_c'  + \ol{\Gvf_c'} -2\Gb) \ol{g'} dm =0.
$$
where
\beq\label{Ggamma}
\Gg:= \frac{2\mu_c \left( 1 - \frac{\Gb}{\Gk_s} \right)}{2(\Gk_m-\Gb)} = \frac{2\mu_c \left( 1 - \frac{\Ga}{2} \right)}{2(\Gk_m-\Gb)}.
\eeq
So by the same argument as above, we infer that the function
$$
\ol{ \left( k_c \Gvf_c- z \ol{\Gvf_c'} -\frac{2\Gb \mu_c}{\Gk_s}z \right)}
- \Gg \ol{(\Gvf_c + z \ol{\Gvf_c'} -2\Gb z)}
$$
can be continued analytically to $D$, namely, there is an analytic function $\eta_2$ in $D$ such that
$$
\left( k_c \Gvf_c- z \ol{\Gvf_c'} -\frac{2\Gb \mu_c}{\Gk_s}z \right)
-\Gg (\Gvf_c + z \ol{\Gvf_c'} -2\Gb z) = \ol{\eta_2} \quad \text{on } \p D ,
$$
which can be rephrased as
\beq\label{etatwo}
\frac{k_c-\Gg}{1+\Gg} \Gvf_c- z \ol{\Gvf_c'}
 = (1+\Gg)^{-1} \ol{\eta_2} + \Gd z \quad \text{on } \p D ,
\eeq
for some real constant $\Gd$. Observe that if $f$ is analytic in $D$, then $\ol{f}$ is a solution (in the complex representation) to the Lam\'e system for any shear modulus $\Gm>0$ and bulk modulus $\Gk>0$. So, $(1+\Gg)^{-1} \ol{\eta_2} + \Gd z$ is a solution to any Lam\'e system. We claim (leaving the proof to the end of this proof) that
\beq\label{largerone}
k_*:= \frac{k_c-\Gg}{1+\Gg} >1.
\eeq
It implies that $k_* \Gvf_c- z \ol{\Gvf_c'}$ is a solution to the Lam\'e system with the shear modulus $\Gm=1$ and the bulk modulus $2 (k_*-1)^{-1}$. So, it follows from \eqnref{etatwo} and uniqueness of the Dirichlet boundary value problem for the Lam\'e system that
$$
k_* \Gvf_c- z \ol{\Gvf_c'}
 = (1+\Gg)^{-1} \ol{\eta_2} + \Gd z \quad \text{in }  D .
$$
We then see that $\Gvf_c'$ is (real) constant in $D$. We also see from \eqnref{etaone} that $\psi_c$ is constant in $D$. In fact, we have from \eqnref{etaone} that
$$
\ol{\psi_c} = \eta_1 + cz \quad\text{on } \p D
$$
for some constant $c$. Since $\psi_c$ and $\eta_1 + cz$ are analytic in $D$, it implies that they are constant in $D$.

Let us now prove \eqnref{largerone}. We see easily from \eqnref{Galpha}, \eqnref{kmbeta} and \eqnref{Ggamma} that
$$
\Gg= \frac{\Gm_c}{2\Gk_s + \Gm_s}.
$$
So we have
$$
(k_c-\Gg)-(1+\Gg) = 2 \Gm_c \left( \frac{1}{\Gk_c} - \frac{1}{2\Gk_s + \Gm_s} \right).
$$
Then \eqnref{largerone} follows by the third condition in \eqnref{hypo}. This completes the proof. \qed

\medskip

\noindent{\sl Proof of Theorem \ref{thm:bulk}}. According to Proposition \ref{prop:shell} (ii), there is $\chi$ such that $\Bu=\nabla \chi$ in $\GO \setminus D$. Since $\Bu=\Bx$ on $\p\GO$, $\Bu=a\Bx+\Bb$ on $\p D$ by \eqnref{axb}, and $\div \Bu$ is constant in $\GO \setminus \ol{D}$, $\chi$ is a solution of the following over-determined problem:
\beq\label{over}
\begin{cases}
\GD \chi = \text{constant} \quad &\text{in } \GO \setminus \ol{D}, \\
\nabla \chi = \Bx &\text{on } \p\GO, \\
\nabla \chi = a\Bx +\Bb &\text{on } \p D.
\end{cases}
\eeq
It is proved in \cite{KL} (see also \cite{KLS}) that if the problem \eqnref{over} admits a solution if and only if $\GO$ and $D$ are concentric disks. This completes the proof. \qed

\section*{Conclusion}
In this paper we prove that if a coated inclusion in two dimensions is neutral to a bulk field, the core and the shell are concentric disks,  provided that the assumption \eqnref{hypo} on elastic moduli holds. It is not clear whether or not there is a coated structure neutral to shear fields, and it is of interest to clarify this. The shear field is the gradient of $\Bh(\Bx)= \BA \Bx$ where $\BA$ is a symmetric matrix whose trace is zero. An extension to three dimensions is also interesting. One can show by the same proof that Proposition \ref{prop:shell} holds to be true in three dimensions. But, we do not know how to prove Proposition \ref{prop:core} in three dimensions.

\section*{Acknowledgement}
The author would like to thank Daewon Chung and Hyundae Lee for helpful discussions.


\begin{thebibliography}{00}

\bibitem{book2} H. Ammari and H. Kang, \textsl{Polarization and moment
tensors with applications to inverse problems and effective medium
theory}, Applied Mathematical Sciences, Vol. 162, Springer-Verlag,
New York, 2007.

\bibitem{AKLL1} H. Ammari, H. Kang, H. Lee, and M. Lim, Enhancement of near cloaking using generalized polarization tensors vanishing structures. Part I: The conductivity problem, Comm. Math. Phys. 317 (2013), 253--266.

\bibitem{CL} R.M. Christensen and K.H. Lo, Solutions for effective shear properties in three phase sphere and cylinder models, J. Mech. Phys. Solids 27 (1979), 315-330.

\bibitem{Chris} R.M. Christensen, A critical evaluation for a class of micromechanics models, J. Mech. Phys. Solids 38 (1990), 379-404.

\bibitem{glu} A. Greenleaf, M. Lassas, and G. Uhlmann, On nonuniqueness for Calderon's inverse problem,
Math. Res. Lett. 10 (2003), 685--693.

\bibitem{hashine} Z. Hashin, The elastic moduli of heterogeneous materials, J. Appl. Mech. 29 (1962), 143--150.

\bibitem{hashine2} Z. Hashin, On elastic behavior of fibre reinforced materials of arbitrary transverse phase geometry, J. Mech. Phys. Solids 13 (1965), 119-134.

\bibitem{HR} Z. Hashin and B.W. Rosen, The elastic moduli fiber-reinforced materials, J. Appl. Mech. 31 (1964), 223--232.


\bibitem{JM} P. Jarczyk and V. Mityushev, Neutral coated inclusions of finite conductivity, Proc. R. Soc. A 468 (2012), 954--970.

\bibitem{KL} H. Kang and H. Lee, Coated inclusions of finite conductivity neutral to multiple fields in two dimensions, Euro. J. Appl. Math., 25 (3) (2014), 329--338.

\bibitem{KLS} H. Kang, H. Lee and S. Sakaguchi, An over-determined boundary value problem arising from neutrally coated inclusions in three dimensions, Annali della Scuola Normale Superiore di Pisa, Classe di Scienze, to appear, arXiv 1501.07465.

\bibitem{KSVW-IP-08} R. V. Kohn, H. Shen, M. S. Vogelius and M. I. Weinstein,
Cloaking via change of variables in electric impedance tomography,
Inverse Problems, 24 (2008), article 015016.

\bibitem{milton}  G.W. Milton,
\newblock {\sl  The Theory of Composites},
\newblock Cambridge Monographs on Applied and Computational
Mathematics, Cambridge University Press, 2001.

\bibitem{MS} G. W. Milton and S. K. Serkov, Neutral coated inclusions in conductivity
and anti-plane elasticity, Proc. R. Soc. Lond. A 457 (2001), 1973--1997.

\bibitem{musk} N.I. Muskhelishvili, {\sl Some Basic Problems of the Mathematical Theory of Elasticity}, English translation, Noordhoff International Publishing, Leyden, 1977.

\bibitem{pendry} J. B. Pendry, D. Schurig and D. R. Smith, Controlling electromagnetic fields, Science 312
(2006), 1780--1782.

\end{thebibliography}
\end{document}